\documentclass{aa}
\usepackage{graphicx}
\begin{document}

\title{The secondary minimum in YY Her: Evidence for a tidally distorted giant}

\subtitle{}

\author{J. Miko\l ajewska\inst{1}, E.A. Kolotilov\inst{2}, 
S.Yu. Shugarov\inst{2,3}, and B.F. Yudin\inst{2,3}}

\offprints{J. Miko\l ajewska, e-mail: mikolaj@camk.edu.pl}

\institute{N. Copernicus Astronomical Center, Bartycka 18, PL-00716, Warsaw, Poland
\and
  Sternberg Astronomical Institute, Universitetskii pr. 13, Moscow, 119899, Russia
\and
  Isaac Newton Institute of Chile, Moscow Branch, Russia}   

\date{Received  / Accepted }

\abstract{We present and analyze quiescent $UBVRI$ light curves of the classical symbiotic 
binary YY Her. We show that the secondary minimum, which is clearly visible only
in the quiescent $VRI$ light curves, is due to ellipsoidal variability of the red giant
component. Our simple light curve analysis, by fitting of the Fourier cosine series,
resulted in a self-consistent phenomenological model of YY Her, in which the 
periodic changes can be described by a combination  of the ellipsoidal changes
and a sinusoidal changes of the nebular continuum and line emission.
}

\titlerunning{Tidally distorted giant in YY Herculis}
\authorrunning{J. Miko{\l}ajewska et al.}

\maketitle

\section{Introduction}

Symbiotic stars are long-period interacting binaries made up of an evolved red giant and a 
hot companion surrounded by an ionized nebula. The hot component in the vast majority of 
systems seems to be a luminous ($\sim 1000\, \mathrm{L_{\sun}}$) and hot ($\sim 10^5\, 
\mathrm{K}$) white dwarf powered by thermonuclear burning of the material accreted from 
its companion's wind. Depending on the accretion rate, these systems can be either in a 
steady burning configuration or undergo hydrogen shell flashes, which in many cases can 
last for decades and even centuries due to low mass of the white dwarf. The latter can be 
the case for non-eruptive symbiotics (such as RW Hya or V443 Her), in which the hot 
components maintain roughly constant luminosity and temperature for many decades 
(Miko{\l}ajewska  \cite{mik97}).  In many systems, however, the hot components show 
multiple outburst activity with timescales of a few/several years, which cannot be simply 
accounted for by the thermonuclear models. Among them, Z And is one of  the best studied 
(e.g. Miko{\l}ajewska \& Kenyon \cite{mk96}, and references therein), and the symbiotic 
stars with multiple outburst activity are often referred to as Z And-type systems.

The photometric history of YY Her since 1890 has been studied in detail by Munari et al. 
(\cite{munari97a}, hereafer M97a) who revealed, in addition to Z And-type multiple 
outburst activity, a periodic fluctuations with $P=590^\mathrm{d}$ and a visual amplitude 
of $\la 0.3^\mathrm{m}$. The orbital nature of this periodicity has been confirmed by 
subsequent UBV photometric observations (Munari et al. \cite{munari97b}; Tatarnikova et 
al. \cite{tatar01}, hereafter M97b and T01, respectively). In particular, the photometric 
and spectroscopic study of YY Her during  the major outburst in 1993  and its decline 
(Tatarnikova et al. \cite{tatar00}, hereafer T00; T01) has shown that the minima are due 
to obscuration of the hot component and the ionized nebula. The return to quiescence in 
1997/98 was accompanied by significant changes in the shape of light curves, and in 
particular by the appearance of a secondary minumum in the $VRI$ light curves (T01; Hric 
et al. \cite{hric01}, hereafter H01). T01 interpreted the secondary minimum in terms of 
ellipsoidal changes of the red giant, whereas H01 argued that the secondary minimum is 
caused by an eclipse of the red giant by a circumstellar envelope  around the hot 
component.

In this paper we present new $UBVRI$ photometric observations obtained during the 
secondary minimum in 2001, and demonstrate that the quiescent light curves of YY Her can 
be succesfully reproduced by a combination of ellipsoidal changes of the cool giant and 
sinusoidal changes of the nebular continuum and line emission.

\section{Observations}

\begin{table}
\caption[]{$UBV$ photometry of YY Her}
\label{table1}
\begin{tabular}[bottom]{lccc}
\hline
\noalign{\smallskip}
JD(2440000+) &  U & B &  V\\
\noalign{\smallskip}
\hline
12047 &  13.73 & 14.24 & 12.95\\
12048 &  13.66 & 14.21 & 12.97\\
12055 &            & 14.26 & 13.10\\
12106 & 13.66  & 14.30 & 13.16\\
12115 & 13.70  & 14.32 & 13.22\\
12135 & 13.38  & 14.28 & 13.32\\
12138 & 13.62  & 14.42 & 13.39\\
12168 & 13.38  & 14.27 & 13.18\\
12172 & 13.46  & 14.20 & 13.14\\
12191 & 13.43  & 14.17 & 13.06\\
12195 & 13.50  & 14.13 & 13.04\\
12202 & 13.38  & 14.13 & 13.01\\
12212 & 13.33  & 14.07 & 12.96\\
12223 & 13.35  & 14.03 & 12.86\\
\noalign{\smallskip}
\hline
\end{tabular}
\end{table}

\begin{table}
\caption[]{$BVR'I'$ photometry of YY Her}
\label{table2}
\begin{tabular}[bottom]{lcccc}
\hline
\noalign{\smallskip}
JD(2440000+) &   B &  V &  R' &  I'\\
\noalign{\smallskip}
\hline
11865 &        & 13.59 & 11.76 & 10.06\\
11866 &        & 13.59 & 11.74 & 10.05\\
11867 &        & 13.64 & 11.76 & 10.09\\
11868 &        & 13.64 & 11.73 & 10.08\\
11887 & 14.87 & 13.40 & 11.58 & 10.00\\
11888 & 14.86 & 13.40 & 11.59 & 10.02\\
11952 &           &13.14 & 11.54 & 10.10\\
12029 &        & 12.91& 11.35 & 9.91\\
12033 &        & 12.94 & 11.33 & 9.90\\
12090 &        & 12.99 & 11.33 & 9.93\\
12162 & 14.27 & 13.14 & 11.51 & 10.08\\
12173 & 14.26 & 13.17 & 11.46 & 10.05\\
12176 &        & 13.08 & 11.43 & 10.03\\
12178 &        & 13.08 & 11.42 & 10.01\\
12188 &        & 13.02 & 11.39 & 9.99\\
12191 & 14.21& 13.02 & 11.36 & 9.97\\
12247 & 14.04 & 12.95 & 11.33 & 9.93\\
\noalign{\smallskip}
\hline
\end{tabular}
\end{table}

\begin{figure}\centering
\resizebox{\hsize}{!}{\includegraphics{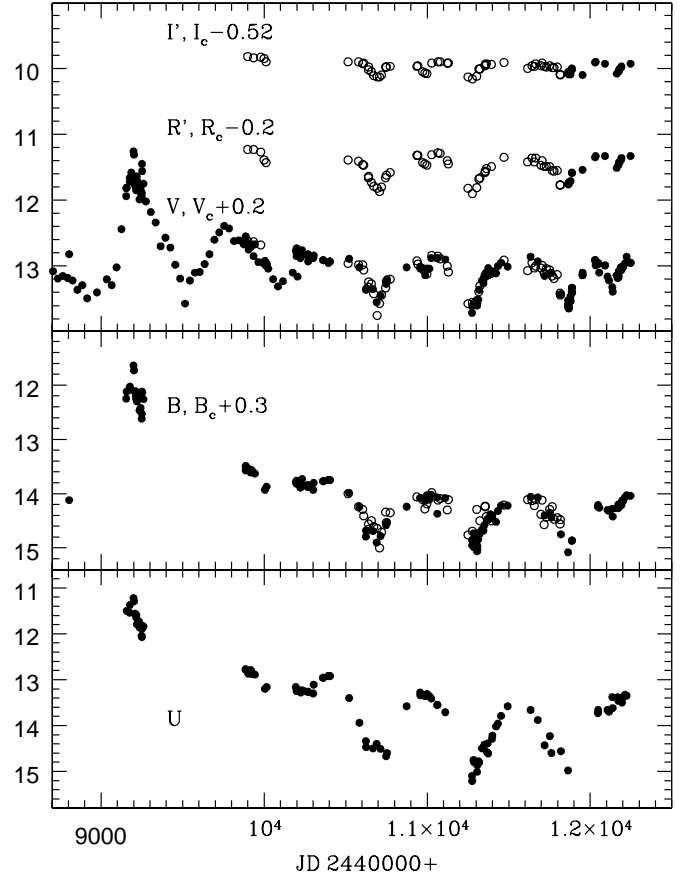}}
\caption{The $UBVRI$ photometry of YY Her in 1993-2001. Closed symbols represent 
our $UBVR'I'$ data (this study; M97a; T00; T01),
and open symbols the $B_\mathrm{c}V_\mathrm{c}R_\mathrm{c}I_\mathrm{c}$ data 
of H01) transformed to our system (see text). }
\label{lc}
 \end{figure}

$UBVR'I'$ photometry was obtained  with the 0.6-m and 1.25-m telescopes at the Crimean 
Station of the Sternberg Astronomical Institute (SAI). For $UBV$ bands 
(Table~\ref{table1}) the 2- channel photometer with an S-20 photomultiplier was used, 
wheras a CCD matrix was the detector for $BVR'I'$ bands (Table~\ref{table2}; see below for 
the definition of $R'I'$). The uncertainty on individual measurements does not exceed 0.03 
mag. The $UBV$ bands correspond to Johnson's broad-band system. The instrumental $ri$ 
bands have the effective wavelengths, $\lambda_\mathrm{r}\approx 7000\, \rm \AA$ and  
$\lambda_\mathrm{i} \approx 9000\, \rm \AA$, and the band widths $\Delta\lambda_\mathrm{r} 
\approx 2400\, \rm \AA$ and  $\Delta \lambda_\mathrm{i} \approx 3000\, \rm \AA$, 
respectively, and they are different  from  any commonly used photometric system. The $ri$ 
observations were reduced to Johnson's $RI$ magnitudes using the standard stars from M67 
(Mendoza, \cite{mendoza67}). In the case of stars with a combination spectra such as YY 
Her, a transformation of the instrumental magnitudes to the standard system may be not as 
precise as in the case of normal stars. For this reason, we refer to these magnitudes as 
$R'I'$. HD 168957 ($U=6.^\mathrm{m}35$, $B=6.^\mathrm{m}91$, $V=7.^\mathrm{m}01$), and the 
star designated as G  ($V=13.^\mathrm{m}08$, $R'=12.^\mathrm{m}44$, $I'=11.^\mathrm{m}94$) 
on the finding chart in M97a) were adopted as the comparison stars.

Our data have been combined with published $UBV$ photometry (T01, and references therein) 
as well as $B_\mathrm{c}V_\mathrm{c}R_\mathrm{c}I_\mathrm{c}$ one performed in the 
modified Johnson-Kron-Cousins system by  H01). The resulting light curves are shown in 
Fig.~\ref{lc}. The $B_\mathrm{c}V_\mathrm{c}R_\mathrm{c}I_\mathrm{c}$ magnitudes in 
Fig.~\ref{lc} were transformed to our system assuming $B=B_\mathrm{c}+0.3$, 
$V=V_\mathrm{c}+0.2$, $R'=R_\mathrm{c}-0.2$, and $I'=I_\mathrm{c}-0.52$, respectively.

\section{Analysis}

The most prominent features  of the $UBV$ light curves of YY Her (Fig.~\ref{lc})
are deep eclipses and the large outburst that started in 1993. 
The $R_\mathrm{c}I_\mathrm{c}/R'I'$ light curves cover the late decline and return to 
quiecence, and they show the characteristic double hump. The secondary minimum is 
also clearly visible in the quiescent $V$ light curve as well as the most recent $B$ data.

\begin{figure}\centering
\resizebox{\hsize}{!}{\includegraphics{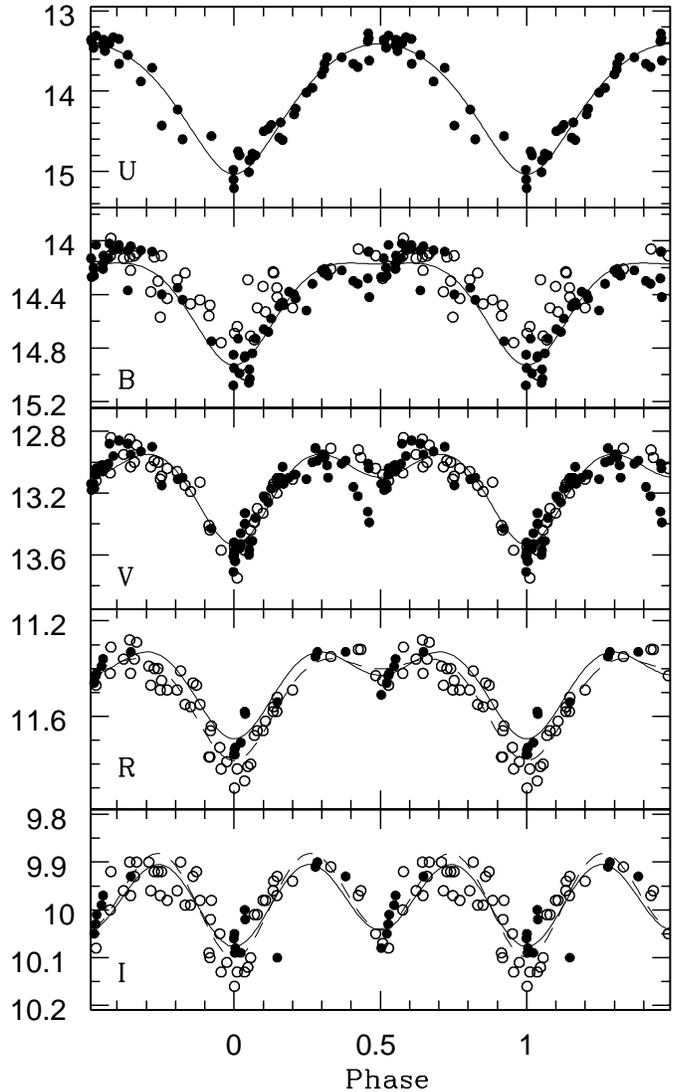}}
\caption{Phased quiescent  light curves of YY Her. The same symbols as in Fig.~\ref{lc}
are used. Our best fit models (Table~\ref{fits}) are plotted as solid lines (our 
data) and dashed lines (H01 data), respectively.}
\label{lc2} \end{figure}

In our analysis, we will focus on the quiescent  light curves. 
Fig.~\ref{lc2} shows the $UBVRI$ magnitudes folded over the last three orbital periods.
We have adopted the ephemeris,
\begin{equation}
\mathrm{Min} = \mathrm{JD}\,2\,450\,686.2(\pm 2.1) + 589.5 (\pm 0.3) \times E,
\end{equation}
derived from 13 individual minima estimates (by a parabolic fit) in published $VRI$ light curves (M97a; M97b; T00; T01; H01). 
The salient characteristic of the quiescent light curves, namely: 
(i) nearly sinusoidal shape of the $U$ light curves; (ii) a flat toped $B$ light curve;
and (iii) the two minima of almost comparable depth in $I'$ light, combined with 
the fact that the quiescent, ultraviolet + optical,  spectra of YY Her can be 
satisfactorily matched with a three-component model composed of a cool giant dominating 
the red spectral range, a gaseous nebula dominating the near UV range, and the hot 
component predominant in the short UV flux (T00), led us to a simple phenomenological 
model in which the light changes are described by a combination of variable nebular 
emission and ellipsoidal changes of the red giant.

Each light curve has been converted to fluxes, $F_\lambda = 10^{-0.4 m_{\lambda}}$, and 
then independently fitted by the Fourier cosine series 
\begin{equation} 
F_{\lambda} = F_{\lambda,0} - A_\mathrm{n,\lambda} \cos \phi - A_\mathrm{g,\lambda} \cos 2 
\phi. 
\end{equation} 
The results of our fits  to the $UBVR'I'$ (our data) and 
$R_\mathrm{c}I_\mathrm{c}$ (H01 data) light curves
are summarized in Table~\ref{fits}, 
columns 2-4. In the case of $U$ data, the best fit is obtained with the $\cos \phi$ term 
only. This indicates that the contribution of the giant to the $U$ light is negligible, 
and that all changes can be attributted to the nebular emission. Assuming that the 
amplitude of the nebular emission changes is the same at all wavelenghts, 
$(F_\mathrm{n,\lambda}^0+A_\mathrm{n,\lambda})/(F_\mathrm{n,\lambda}^0- 
A_\mathrm{n,\lambda}) \approx (F_\mathrm{U,0}+A_\mathrm{U})/(F_\mathrm{U,0}- 
A_\mathrm{U})=C \sim 4.5$, we can separate the contributions of the nebular and the giant 
components to  the constant $F_{\lambda,0}$. The resulting charateristics of the nebular 
source (columns 5--7) and the red giant (columns 8--10) are also given in  
Table~\ref{fits}. 

The maximum, $F_\mathrm{\lambda,max} (\phi=0.5) = F_\mathrm{n,\lambda}^0+A_\mathrm{n,\lambda}$,  fluxes 
of the nebular component have been calculated using the zeroth magnitude star
fluxes from Strai\v{z}ys (\cite{straizys92}). The $UBVR'I'$ and $R_\mathrm{c}I_\mathrm{c}$  magnitudes
corresponding to these maximum fluxes have been also calculated. We note here that the broad-band magnitudes,
and the resulting fluxes may be significantly afftected by emission lines. Using the 
quiescent optical/red spectra of YY Her  (T00; T01), we have calculated the contribution 
of the emission lines to the observed $BVR'$ magnitudes as $\Delta m \ga 0.15/0.02/0.08$ 
mag, respectively (this contribution can be, in fact, even higher because we have not 
included faint lines). Using these estimates, we have  found that the nebular line 
emission is responsible for at least 30\,\%($B$ band), 7\,\%($V$) and  50\,\%($R'$) of the 
total nebular emission fluxes, and the corresponding magnitudes,  given in the 5th and 6th 
columns of Table~\ref{fits}. Finally, the maximum emission  measure of the nebular 
continuum (7th column of Table~\ref{fits}) has been estimated from the maximum, 
$F_\mathrm{\lambda,max}$, fluxes of the nebular component corrected for the contribution 
from emission lines and the interstellar reddening $E_\mathrm{B-V} = 0.2$ (M97b),

\begin{equation}
EM=n_\mathrm{e}^2 V/4\pi d^2=F_\mathrm{\lambda,max}/\epsilon_\lambda^\mathrm{bf+ff}.
\end{equation}
The volume emission coefficients, $\epsilon_\lambda^\mathrm{bf+ff}$, have been estimated
using our $UBVR'I'$ filter transmission functions and bf+ff continuum emission coefficients (case B recombination) from Pottash (\cite{pottash84}). 
The $EM$ values derived for each band agree with each other to within a factor of 2, and they cluster around $EM \sim 2\times 10^{14}$ and $\sim 3\times 10^{14} \mathrm{cm}^{-5}$ for $T_\mathrm{e} = 10^4$ and $2\times10^4\,\mathrm{K}$, respectively.

The nebular emission measure derived from our light curve analysis requires a Lyman 
continuum luminosity of $\sim 6$--$5 \times 10^{45}\, 
(d/1\,\mathrm{kpc})^2\,\mathrm{phot\,s^{-1}}$ for $T_\mathrm{e} = 10^4$--$2\times10^4\, 
\mathrm{K}$, which corresponds to a hot component with $T_\mathrm{h} \sim 10^5\, 
\mathrm{K}$ and $L_\mathrm{h} \sim 80 (d/1\,\mathrm{kpc})^2 \mathrm{L}_{\sun}$. This can 
be compared with independent estimates  for the quiescent $L_\mathrm{h}$. In particular, 
the bolometric fluxes for the hot component given by M97b and T00 correspond to 
$L_\mathrm{h} \sim (30$--$60)\, (d/1\,\mathrm{kpc})^2 \mathrm{L}_{\sun}$. 

The $UBVR'I'$ and $R_\mathrm{c}I_\mathrm{c}$  magnitudes of the cool component have been 
calculated using $m_0=-2.5 \log F_\mathrm{g,\lambda}^0$, 
$m_\mathrm{max}=-2.5 \log  (F_\mathrm{g,\lambda}^0+A_\mathrm{g,\lambda})$, and 
$m_\mathrm{min}=-2.5 \log  (F_\mathrm{g,\lambda}^0-A_\mathrm{g,\lambda})$, where 
$F_\mathrm{g,\lambda}^0$ is the cool component contribution to the constant 
$F_\mathrm{\lambda,0}$. The colours derived from 
these magnitudes, $B-V=1.75/1.77/1.73$, $V-R'=1.81/1.73/1.92$ and $V-I'=3.45/3.33/3.61$ 
representing the average, maximum and minimum values, respectively, are in excellent 
agreement with those expected for a moderately reddened M3--M4 giant. Moreover, they 
indicate somewhat higher temperature (earlier spectral type) at maxima ($\phi=0.25$ and 
0.75) than that at minima ($\phi=0$ and 0.5) as expected for a tidally distorted giant.

Assuming the Roche-lobe filling factor $R_\mathrm{g}/R_\mathrm{L} \approx 1$, the 
gravity-darkening exponent, $\alpha = 0.32$ (convective envelope), 
and the system inclination $i \sim 70$--$90^{\degr}$, the amplitude of the cool component 
changes in $I'$ is consistent 
with a mass ratio $q=M_\mathrm{g}/M_\mathrm{h} \sim 2$--3. This estimate is based on 
synthetic light curves calculated using the Wilson-Devinney (WD) code (see Belczy{\'n}ski 
\& Miko{\l}ajewska \cite{bm98} for details). The amplitudes in $R,V$ bands, however, 
require a much higher value of $\alpha \ga 1$. Similar effect was found in T CrB, and it 
is most likely due to the blackbody approximation for wavelength dependence in the WD code 
used to calculate the synthetic light curves (Belczy{\'n}ski \& Miko{\l}ajewska 
\cite{bm98}). This assumption is probably not valid in the case of M-type stars with 
strong contribution of TiO bands in the optical and red part of the spectrum. Recently, 
Orosz \& Hauschildt (\cite{oh00}) have demonstrated that, in general, the light curves for 
tidally distorted cool giants computed using the NEXTGEN model atmospheres for cool giants 
have deeper minima in the optical range then their blackbody counterparts.  

Finally, we note that the amplitude of ellipsoidal variability is a strong function of the 
Roche-lobe filling factor, $\Delta\,m \propto (R_\mathrm{g}/R_\mathrm{L})^3$ (e.g. Hall 
\cite{hall90}). The lower limit for the cool component changes set by the uncertainty of 
our fits (Table~\ref{fits}) then requires $R_\mathrm{g}/R_\mathrm{L} \ga 0.9$ for $q \ga 
1$ and $i \ga 70^{\degr}$.

\begin{table*}
\caption[]{The best fit parameters for YY Her}
\label{fits}
\begin{tabular}[bottom]{cccccccccc}
\hline
\noalign{\smallskip}
\multicolumn{1}{c}{Filter} & \multicolumn{3}{c}{Best fit parameters} &\multicolumn{3}{c}{Hot nebular source} &\multicolumn{3}{c}{Cool giant}\\
    & $F_\mathrm{\lambda,0}^{(1)}$ & $A_\mathrm{n,\lambda}^{(1)}$ & $A_\mathrm{g,\lambda}^{(1)}$ &
 $F_\mathrm{\lambda,max}^{(2)}$ & $m_\mathrm{\max}$\,[mag] &$EM^{(3)}$ &$m_0$\,[mag] & $m_\mathrm{max}$\,[mag] & $m_\mathrm{min}$\,[mag] \\
\noalign{\smallskip}
\hline
$U$ & $2.65\pm0.05$ & $1.68\pm0.06$ & 0 & $17.8\pm0.4$ & $13.41\pm0.03$ & 1.5/2.7 & & & \\
$B$ & $1.77\pm0.02$ & $0.54\pm0.03$ & $0.17\pm0.03$ & $8.6\pm0.3$ & $14.64\pm0.06$ & 3.0/2.4 &$15.21\pm0.06$ & $15.03\pm0.08$ & $15.43\pm0.11$\\
$V$ & $5.64\pm0.05$ & $0.96\pm0.06$ & $0.84\pm0.07$ &
$9.0\pm0.4$ & $14.02\pm0.06$ & 2.8/3.2 & $13.46\pm0.04$ & $13.26\pm0.05$ & $13.70\pm0.07$\\
$R'^{(4)}$ & $26.4\pm0.4$ & $2.9\pm0.4$ & $2.5\pm0.4$ &
$12.4\pm0.9$ & $12.82\pm0.14$ & 2.0/2.9 & $11.65\pm0.03$ & $11.53\pm0.04$ & $11.78\pm0.05$\\
$R_\mathrm{c}^{(5)}$ & $26.5\pm0.2$ & $3.4\pm0.3$ & $1.7\pm0.3$
& $19.7\pm0.9$ & $12.64\pm0.09$ & 3.2/4.5 & $12.00\pm0.03$ & $11.89\pm0.04$ & $12.13\pm0.05$\\
$I'^{(4)}$ & $101.9\pm0.9$ & $1.5\pm0.7$ & $7.2\pm1.1$ & 
$3.0\pm0.9$ & $13.5\pm0.5$ & 1.6/2.0 & $10.01\pm0.02$ &$9.93\pm0.03$ & $10.09\pm0.04$\\
$I_\mathrm{c}^{(5)}$ & $63.5\pm0.6$ & $1.6\pm0.8$ & $5.4\pm0.8$ &
$4.9\pm1.5$ & $13.5\pm0.5$ & 2.5/3.0 & $10.54\pm0.03$ &
$10.44\pm0.04$ & $10.64\pm0.05$\\
\noalign{\smallskip}
\hline
\end{tabular}
(1) in units of $10^{-6}$; 
(2) in units of $10^{-15}\, \mathrm{erg\,s^{-1}\,cm^{-2}\AA^{-1}}$; 
(3) maximum emission measure in units of $10^{14} \mathrm{cm}^{-5}$, for $T_\mathrm{e} = 10^4/2\times10^4\, \mathrm{K}$; 
(4)  our $R'/I'$ data; (5) $R_\mathrm{c}/I_\mathrm{c}$ from H01).
\end{table*}

The presence of a Roche-lobe filling giant provides an opportunity to estimate the 
distance $d$ to YY Her. The Roche lobe radius is $R_\mathrm{L}/a \approx 0.44\pm0.05$ for 
the mass ratio of $q=M_\mathrm{g}/M_\mathrm{h} \sim 1$--3 (Paczy{\'n}ski \cite{bep71}). 
Our orbital period requires the binary separation, $a\approx1.85\pm0.15\, \mathrm{a.u.}$ 
for a reasonable total mass of the system, $M_\mathrm{g}+M_\mathrm{h} \sim 2$--$3\, 
\mathrm{M}_{\sun}$ (e.g. Miko{\l}ajewska \cite{mik97}), so $R_\mathrm{L} \approx 0.8 \pm 
0.2\, \mathrm{a.u.}$ Using the Barnes-Evans relation (Cahn \cite{cahn80}),
\begin{equation}
F_\mathrm{K} = 4.2207 -0.1 K - 0.5 \log s,
\end{equation}
with the $K$ surface brightness, $F_\mathrm{K} = 3.84$ (M4\,III), and $K_0=7.9$,
we estimate the giant's angular diameter, $s=0.152\, \mathrm{mas}$, and $d = 10\pm3\, 
\mathrm{kpc}$. 
Such a distance implies a very high $\sim 3\, \mathrm{kpc}$ height above
galactic plane  and locates YY Her  in galactic halo.

\section{Concluding remarks}

Our very simplistic light curve analysis, namely independent fitting of the Fourier cosine 
series to the quiescent $UBVR'I'$ light curves, resulted in a self-consistent 
phenomenological model of YY Her, in which the periodic changes can be described by 
variations of the nebular emission along the orbit (the $\cos \phi$ term) combined with 
the ellipsoidal variability of the red giant (the $\cos 2\phi$ term).

YY Her is not the only symbiotic system with the secondary minima. Similar light curve 
behaviour is observed in CI Cyg, whose quiescent $UBVRI$ light curves (Miko{\l}ajewska 
\cite{mik01}) are almost identical with the light curves shown in Fig.~\ref{lc2}. The 
secondary minima are also present in quiescent near infrared light curves of BF Cyg 
(Miko{\l}ajewska et al. \cite{mik02}).  

We also think that the alternative interpretation of the secondary minimum in terms of the 
eclipse of the cool giant by the circumstellar envelope of the hot component (H01) is 
implausible. First, it cannot be the ionized nebular envelope because to cause eclipses it 
must be optically thick in the optical/red continuum whereas the presence of a prominent 
Balmer jump indicates that it is not. Second, to account for the amplitudes of the 
secondary minimum, it must screen $\sim 20$\,\% of the red giant's projected surface. Such 
a big and hot screen, however, should dominate the quiescent optical continuum, which is 
not observed. Finally, we can also exclude eclipses of the cool giant by an accretion 
disc. A geometrically and optically thick disc seen nearly edge-on could mimic a cool 
source, and screen significant part of the giant's surface. Such a disc, however, would 
also completely hide the hot white dwarf wheras the quiescent shortwavelength {\it IUE} 
spectrum clearly shows a hot continuum (M97b). Similarly, a hot rising continuum is 
present in quiescent shortwavelength {\it IUE} spectra of CI Cyg and BF Cyg.

One of fundamental questions in relation of symbiotic binaries is the mechanism that 
powers the Z And-type multiple outburst activity observed in many classical symbiotic 
systems. The quiescent spectra of these systems can mostly be fitted by a hot stellar 
source (most likely a white dwarf powered by thermonuclear burning of the material 
accreted from its companion) and its ionizing effect on a nebula (e.g. M{\"u}rset et al. 
\cite{murset91}; M97b, Miko{\l}ajewska \& Kenyon \cite{mk96}). Their outburst activity, 
however, with time scales of a few/several years cannot be simply accounted by the 
thermonuclear models. A promising interpretation of this activity involves changes in mass 
transfer and/or accretion disc instabilities. Detection of an ellipsoidal hot continuum 
source during outbursts of CI Cyg, AX Per, YY Her, AS 338 and other Z And-type systems 
suggests the presence of an optically thick accretion disc (Miko{\l}ajewska \& Kenyon 
\cite{mk92}; Esipov et al. \cite{esipov00}; T00; Miko{\l}ajewska et al. \cite{mik02}) 
strongly supports this interpretation.

Related to this problem is the process of mass transfer -- Roche lobe overflow or stellar 
wind -- and the possibility of an accretion disc formation. The presence of secondary 
minima in the light curves of YY Her, apparently due to ellipsoidal changes of the red 
giant, as well as in a few other similar systems provides important clues. Although it is 
very premature to claim that all symbiotic systems with the Z And-type activity do have 
tidally distorted giants, and -- at least during active phase -- accretion discs, whereas 
the non-eruptive systems (such as RW Hya) do not, the former is certainly the case for YY 
Her, CI Cyg and a few other active systems. 

Systematic searches for ellipsoidal variations in both active and non-eruptive symbiotic 
stars are necessary to address the problem, and confirm or exclude tidally distorted 
giants. The observations must be, however, carried in the red and near-IR range where the 
cool giant dominates the continuum light. We also note that any optical/red light curve 
analysis must be supported by spectroscopic information about contamination of the 
broad-band photometry by the nebular component. 

\begin{acknowledgements} 
This research was partly supported by KBN research grant No. 5P03D 019 20,
and RFBR grant Nos. 02-02-16235, 02-02-16462 and 02-02-17524.
\end{acknowledgements}

\end{document}